\documentclass[oneside,10pt]{swissi}

\newcommand{\papertitle}{Firm Valuation When AI Shapes the Business Model}
\newcommand{\papersubtitle}{A Milestone-Based Real-Options Framework for the AI Valuation Uncertainty Problem}

\newcommand{\papercitationauthors}{Kurz, W. and Stricker, W. and Marx, S. and Reinhardt, F. and Kollberg, F.}

\newcommand{\paperauthors}{%
  Walter Kurz\textsuperscript{1}, %
  Wojtek Stricker\textsuperscript{1}, %
  Stefan Marx\textsuperscript{2}, %
  Frank Reinhardt\textsuperscript{2}, %
  Florian Kollberg\textsuperscript{2}%
}

\newcommand{\paperaffiliations}{%
  \textsuperscript{1}Swissi Institute for AI,
  \href{mailto:kurz@swissi-ai.institute}{kurz@swissi-ai.institute};
  \href{mailto:stricker@swissi-ai.institute}{stricker@swissi-ai.institute}\\
  \textsuperscript{2}Hochschule f\"ur Wirtschaft und Umwelt N\"urtingen-Geislingen%
}

\newcommand{\paperdeclarationaffiliations}{%
  \begin{tabular}[t]{@{}l@{}}
    \textsuperscript{1}Swissi Institute for AI,
    \href{mailto:kurz@swissi-ai.institute}{kurz@swissi-ai.institute};
    \href{mailto:stricker@swissi-ai.institute}{stricker@swissi-ai.institute}\\
    \textsuperscript{2}Hochschule f\"ur Wirtschaft und Umwelt N\"urtingen-Geislingen
  \end{tabular}%
}

\newcommand{\papercorresponding}{%
  Walter Kurz,
  \href{mailto:kurz@swissi-ai.institute}{kurz@swissi-ai.institute},
  \href{https://orcid.org/0009-0006-8045-4775}{ORCID 0009-0006-8045-4775}%
}

\newcommand{\paperfunding}{%
  This research received no external funding.%
}
\newcommand{\paperconflictsofinterest}{%
  The authors declare no conflicts of interest.%
}
\newcommand{\paperdataavailability}{%
  The data and code supporting this article are available from the corresponding author on reasonable request.%
}
\newcommand{\paperaitools}{%
  Generative AI tools were used for editorial work only, such as language editing and formatting, in accordance with international scientific standards. The authors verified the content and remain fully responsible for the article.%
}
\newcommand{\paperauthorcontributions}{%
  All authors contributed equally to this article.%
}

\usepackage{xltabular}

\title{\papertitle \\[0.35em]
\normalsize \papersubtitle}

\author{{\fontsize{8}{10}\selectfont\paperauthors}\\[0.9em]
{\scriptsize\paperaffiliations}}
\date{\vspace{1.5em}}

\begin{document}
\maketitle

\begin{abstract}
Standard valuation methods, including discounted cash flow, the income approach standard IDW~S~1 of the Institute of Public Auditors in Germany, and market multiples, compress milestone probabilities, continuation options, and risk shifts into opaque aggregate parameters; none provides a structured protocol for decomposing AI integration into auditable option-level assumptions. We propose an industry-agnostic taxonomy separating AI Integrators from AI Providers. AI Integrators are further classified by their Integration Depth Level, ranging from no integration to AI at the core of the product or process. A milestone-gated real-options overlay decomposes milestone state value into five components, and an Analytic Hierarchy Process-based Success Readiness Index derives per-option probabilities from structured pairwise comparisons for scenario analysis. Applied to an AI-native energy software-as-a-service firm, the framework yields a coherent valuation band traceable to identifiable option-level assumptions. Risk concentrates in later-stage continuation options, matching the structural prediction for AI Providers. The protocol applies across the firm lifecycle, including mergers and acquisitions due diligence. The case is a single-firm demonstration of protocol coherence, not empirical validation; multi-case testing against realised post-exit valuations is left to future research.
\end{abstract}


\section{Introduction}

Artificial intelligence has moved from pilot budgets to strategic capital allocation. In the United States, AI-focused firms have reached private-market valuations of hundreds of billions of dollars within years of founding,\cite{anthropic2026seriesg,databricks2025seriesk} while no European company created in the last fifty years has achieved a market capitalisation above EUR~100~billion.\cite{draghi2024competitiveness} The speed and magnitude of this repricing are not explained by technology alone. Value realisation remains highly heterogeneous across firms, sectors, and jurisdictions,\cite{mckinsey2025rewiring,mckinsey2025stateai,brynjolfsson2023generativework,eisfeldt2023firmvalues} and the pattern is consistent: most deployments remain at limited integration depth, while a smaller group restructures workflows, decision rights, and data feedback loops.\cite{mckinsey2025rewiring,mckinsey2025stateai} The difference between experimenting with AI and restructuring around it is a structural divide, and standard valuation practice has yet to absorb it. The standard discounted cash-flow framework values a firm as
\begin{equation}\label{eq:dcf}
V_0 \;=\; \sum_{t=1}^{T} \frac{\mathit{CF}_t}{(1+r)^t} \;+\; \frac{\mathit{TV}}{(1+r)^T}
\end{equation}
where $\mathit{CF}_t$ denotes expected free cash flow, $r$ the cost of capital, and $\mathit{TV}$ a terminal value capturing perpetual growth.\cite{koller2020valuation} This architecture works when the cash-flow engine is stable and projectable from historical patterns. AI integration strains this architecture beyond its design range: it restructures the engine itself through staged technical, organisational, and regulatory decisions.\cite{schwartz2000internet,benaroch1999isre} Each milestone along the integration path (a workflow redesign, a proprietary model deployment, a regulatory clearance) changes what $\mathit{CF}_t$ will look like downstream. The variable the analyst must forecast is itself being reshaped by decisions the firm has not yet made. An analyst who compresses this into a single growth-rate assumption is implicitly pricing every milestone probability, every continuation option, and every risk shift in one opaque number. The DCF number itself may remain coherent; the reasoning behind it becomes invisible, so two analysts who disagree cannot locate the source of their disagreement.

\begin{figure}[htbp]
\centering
\begin{subfigure}[t]{0.48\textwidth}
\centering
\begin{tikzpicture}[scale=0.9]
  \draw[->, thick] (0,0) -- (7.8,0) node[right, font=\small] {$t$};
  \draw[->, thick] (0,0) -- (0,4.8);

  \node[rotate=90, anchor=south, font=\small] at (-0.6,2.4) {uncertainty ($\sigma$)};

  \draw[dashed, gray] (0,0.3) -- (7.5,0.3);
  \node[left, font=\footnotesize, gray] at (0,0.3) {0};

  \draw[dotted, thick] (2.0,0) -- (2.0,4.5);
  \node[above, font=\footnotesize] at (2.0,4.5) {founding};

  \node[below, font=\footnotesize] at (1.0,-0.1) {pre-seed};
  \node[below, font=\footnotesize] at (4.8,-0.1) {operational};

  \draw[dotted, thick, red!60!black] plot[smooth, tension=0.7]
    coordinates {(0.1,4.0) (0.5,3.95) (1.0,3.85) (1.5,3.75) (2.0,3.6)};
  \draw[thick, red!60!black] plot[smooth, tension=0.7]
    coordinates {(2.0,3.6) (2.8,3.0) (3.5,2.4) (4.5,1.7) (5.5,1.2) (6.5,0.85) (7.3,0.7)};

  \node[above, font=\footnotesize, red!60!black] at (7.3,0.7) {$\sigma_{\mathrm{DCF}}$};
\end{tikzpicture}
\caption{Lifecycle: $\sigma$ before and after founding.}
\label{fig:dcf-panel-a}
\end{subfigure}
\hfill
\begin{subfigure}[t]{0.48\textwidth}
\centering
\begin{tikzpicture}[scale=0.9]
  \draw[->, thick] (0,0) -- (7.8,0) node[right, font=\small] {$t$};
  \draw[->, thick] (0,0) -- (0,4.8);

  \node[rotate=90, anchor=south, font=\small] at (-0.6,2.4) {uncertainty ($\sigma$)};

  \draw[thick, red!50!black, densely dash dot] (2.5,0) -- (2.5,4.5);
  \node[above, font=\footnotesize, red!50!black, align=center] at (2.5,4.5) {AI introduced};

  \draw[dotted, gray] (4.5,0) -- (4.5,4.3);
  \node[above, font=\scriptsize, gray] at (4.5,4.3) {$m_1$};
  \draw[dotted, gray] (5.8,0) -- (5.8,4.3);
  \node[above, font=\scriptsize, gray] at (5.8,4.3) {$m_2$};

  \draw[thick, blue!70!black] plot[smooth, tension=0.7]
    coordinates {(0.2,1.0) (0.8,0.95) (1.5,0.9) (2.0,0.85) (2.5,0.8)};

  \draw[thick, red!60!black] (2.5,0.8) -- (2.5,3.5);

  \draw[thick, red!60!black] plot[smooth, tension=0.5]
    coordinates {(2.5,3.5) (3.0,3.3) (3.5,3.0) (4.0,2.7) (4.5,2.3) (5.2,2.0) (5.8,1.7) (6.5,1.4) (7.3,1.2)};

  \fill[red!8] plot[smooth, tension=0.5]
    coordinates {(2.5,3.5) (3.0,3.3) (3.5,3.0) (4.0,2.7) (4.5,2.3) (5.2,2.0) (5.8,1.7) (6.5,1.4) (7.3,1.2)}
    -- (7.3,0.8) -- (2.5,0.8) -- cycle;

  \draw[dashed, blue!40!black] (2.5,0.8) -- (7.3,0.8);

  \node[below, font=\footnotesize] at (1.3,-0.1) {pre-AI};
  \node[below, font=\footnotesize] at (5.0,-0.1) {integration};

  \node[font=\footnotesize, red!60!black] at (5.0,3.0) {$\sigma_{\mathrm{AI}}$};
  \node[font=\footnotesize, blue!40!black] at (5.5,0.55) {pre-AI $\sigma$};
\end{tikzpicture}
\caption{AI introduction: $\sigma$ spike and milestone-driven decay ($m_1$, $m_2$). Shaded area: additional uncertainty introduced by integration.}
\label{fig:dcf-panel-b}
\end{subfigure}
\caption{Projection uncertainty $\sigma$ over the firm lifecycle~(a) and around AI introduction~(b). Illustrative.}
\label{fig:dcf-uncertainty}
\end{figure}
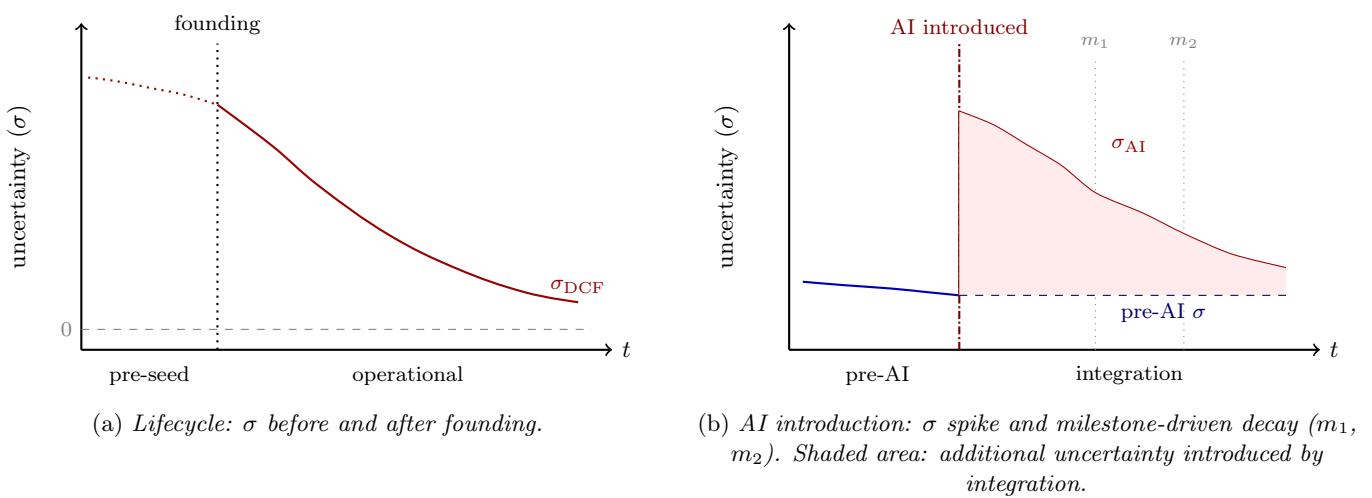

The accuracy of any DCF valuation depends on the data available to the valuator at the valuation date, and that data is a function of operating history. \Cref{fig:dcf-uncertainty} illustrates the valuator's problem. Before founding, there is no observable data: projection uncertainty $\sigma$ is at its maximum. The act of establishing a firm marks a transition; from that point, costs, contracts, and revenue begin to accumulate, and $\sigma$ decreases as the projection gains empirical grounding. The longer the firm operates, the lower $\sigma$ becomes, but it never reaches zero because DCF remains a forecast at every point. When an operational firm introduces AI into its processes, $\sigma$ spikes: the business model is being reconfigured, and the historical patterns the valuator relied on lose predictive power. The spike is not gradual; at sufficient integration depth, it is discontinuous. As AI milestones clear, $\sigma$ decreases again, but the rate of recovery depends on the depth and scope of the integration, and $\sigma$ may not return to its pre-AI level because the business model has structurally changed.

What makes AI integration particularly resistant to ex ante valuation is the confluence of mechanisms it activates. Even within a given integration regime, deployment choices exhibit sensitive dependence on initial conditions: small differences in sequencing produce divergent firm-level trajectories because each decision reshapes the capability base available for subsequent ones.\cite{lorenz1963chaos} The resulting system exhibits non-linear feedback loops across functions, costs, and competitive position, characteristic of complexity-economic regimes where equilibrium assumptions break down and the distribution of outcomes is fat-tailed rather than normally distributed around a central projection.\cite{arthur2021complexity} Strategic interaction amplifies this further: platform economics, winner-take-most dynamics, and moat construction turn each firm's AI decisions into moves in an evolving competitive game whose payoff structure changes as other players act.\cite{vonneumann1944games,rochet2003platform} For a valuator, these mechanisms operate simultaneously. The future configuration of an AI-integrated firm carries two stacked layers of uncertainty: magnitude uncertainty driven by the firm's own decisions, and kind uncertainty driven by what competitors do.

Within DCF, three mechanisms compound the problem. First, staged optionality collapses into the long-term growth rate: a contingent, milestone-gated payoff that management may abandon or expand is priced as if it were a deterministic cash-flow trajectory. Second, risk concentrates in a single discount rate that cannot represent the structural shift from high uncertainty at early integration stages to lower uncertainty after milestone clearance. Third, management flexibility (the right to defer, expand, or abandon at each decision node) carries positive value that DCF assigns implicitly at zero.

U.S. private-market practice already reveals this tension. Venture rounds price preferred claims with contractual protections, producing headline post-money valuations that average roughly 48\% above common-share fair value.\cite{gornall2020squaring} Tax and disclosure workflows value common shares separately under dedicated standards such as 409A and pre-IPO scenario methods,\cite{us26cfr1409a1,maplebear2023s1correspondence} so the same firm carries two structurally different valuations at the same date. The very existence of this dual-track architecture signals that practitioners already distinguish between option-loaded headline figures and fair-value estimates. A European practitioner reading a U.S. post-money figure as a common-equity equivalent makes a structural category error: the figure prices contractual optionality alongside operating forecasts.

In European and Swiss settings, the mismatch runs deeper. Dominant valuation standards such as IDW~S~1 mandate discounted earnings as the primary method and require demonstrable historical patterns.\cite{idw2016s1} EU and Swiss AI governance frameworks formalise compliance gates that affect execution cost, admissibility risk, and time-to-scale.\cite{eu2024aiact,finma2024guidance08,bakom2025aiapproach} These instruments were built for firms with established cash-flow histories, and for those firms they work well. The problem arises when the same instruments are applied to milestone-gated AI ventures: not as a deliberate analytical choice, but because no structured alternative exists within the European valuation ecosystem. The result is a measurement mismatch, not conservative valuation. European venture capital stands at less than one-third of U.S. levels as a share of GDP, and comparable startups are valued approximately seven times higher in the United States at the same development stage.\cite{arnold2024steppingup,draghi2024competitiveness} Part of this gap reflects genuine differences in market structure and risk appetite. Part of it may reflect the absence of a valuation framework that can price what these firms actually are.

The core problem is not whether AI matters for firm value, but how to value integration pathways whose pay-offs depend on depth, sequence, and jurisdiction-conditioned feasibility. Existing literature provides the necessary components: real options for staged technology investment, multi-criteria methods for structured probability assessment, maturity models for depth classification, and market-pricing studies for empirical calibration.\cite{schwartz2000internet,benaroch1999isre,saaty1980ahp,hansen2024aicmm,eisfeldt2023firmvalues} The unresolved task is to integrate these components into a single valuation logic that identifies the relevant milestones, specifies how each probability is estimated, and indicates where integration depth changes the parameter set. Within the present work, a structured, auditable protocol is developed that decomposes AI integration into option-level valuation inputs, comparable across analysts, applicable across integration depths, and consistent with existing valuation standards.

\section{Related Work}

Valuation research is still dominated by discounted cash-flow logic calibrated on relatively stable business models.\cite{koller2020valuation} The young-firm literature shows why that architecture weakens under negative earnings, high failure probability, and limited historical depth: conventional DCF and relative valuation break down precisely where the firm profile is dominated by intangible investment, survival uncertainty, and non-linear growth trajectories.\cite{damodaran2018darkside,damodaran2009young} Intangible-intensive firms compound this problem because their core value drivers resist balance-sheet measurement.\cite{lev2001intangibles,haskel2017capitalism} The unresolved issue is not whether DCF remains useful, but where its assumptions become structurally incomplete for staged AI integration.

Market-based and early-stage methods face structurally different but equally limiting constraints. Revenue and ARR multiples, the dominant approach for private AI firm valuation in practice, compress option value into a comparable figure without identifying which milestone generates the premium or how sensitive the valuation is to individual probability assumptions; their application is also constrained by the scarcity of comparables for firms at Integration Depth Level (IDL)~2--3.\cite{damodaran2009young,eisfeldt2023firmvalues} Early-stage methods such as the VC method and First Chicago introduce scenario weighting but stop short of per-milestone probability estimation and depth-conditioned option architecture.\cite{lerner2020vc} The income approach under IDW~S~1, the primary European statutory method, shares DCF's structural limitation: it projects a single income path and discounts it, compressing milestone probabilities and continuation-option value into the same opaque aggregate parameter.\cite{idw2016s1} The framework developed here does not replace these methods; it adds a decomposition layer in the form of option-level inputs with independent probability estimates linked to identifiable milestone assumptions.

Real-options research addresses staged commitment under uncertainty and supplies the formal language for sequential exercise, deferral, expansion, and abandonment.\cite{myers1977borrowing,dixit1994investment,trigeorgis1996realoptions} Compound-option formulations extend this to multi-stage investment chains where each decision gate conditions the next.\cite{geske1979compound,cassimon2004nfold} Applied work in internet and IT contexts confirms that technology investments are path-dependent and decision-contingent, with value concentrated in future optionality rather than static baseline projections.\cite{schwartz2000internet,schwartz2001rational,benaroch1999isre,benaroch2000misq} This research programme has also extended to natural-resource investment, where empirical evidence confirms that firms hold loss-making assets open when waiting option value exceeds closure value,\cite{moeltufano2002mines} and to strategic investment contexts where option value interacts with competitive pre-emption and deterrence.\cite{smittrigeorgis2004strategic} Across all applied domains, the consistent finding is that option value is substantial relative to static NPV and that standard discounted-earnings methods systematically underestimate flexibility value. These models treat technology generically: none addresses AI integration specifically or distinguishes integration depth as a structural variable that reshapes the option parameters themselves. Stage-level probabilities in applied implementations are typically derived from market-implied volatility or unstructured expert judgement, with no structured protocol for estimating each option's likelihood independently or varying individual probabilities in scenario analysis.

AI adoption and maturity studies document heterogeneous operational effects and depth-dependent capability formation.\cite{hansen2024aicmm,dreyling2024aimm} Productivity evidence shows strong heterogeneity by worker skill and organisational complementarity,\cite{brynjolfsson2023generativework} delayed measured gains from costly intangible co-investments,\cite{brynjolfsson2018jcurve} and market repricing that varies with AI exposure.\cite{eisfeldt2023firmvalues} Industry surveys confirm near-ubiquitous experimentation but limited deep workflow redesign.\cite{mckinsey2025rewiring,mckinsey2025stateai} This stream identifies the empirical phenomenon (integration depth matters) but does not provide a valuation operator that transforms depth and milestone progression into auditable firm-value components.

Existing maturity scales classify depth within individual organisations; none distinguishes firms that embed AI into existing operations from firms whose core product is AI itself. The two cases generate structurally different risk profiles, option architectures, and competitive dynamics. No classification in this literature provides an industry-agnostic taxonomy that maps these distinctions to valuation mechanics.

Multi-criteria decision analysis provides structured tools for aggregating expert judgement under complexity. The Analytic Hierarchy Process decomposes assessment into pairwise comparisons with a built-in consistency diagnostic,\cite{saaty1980ahp} and has been combined with real options for telecom and IT infrastructure investment decisions.\cite{angelou2008ahpro} Alternative methods (the Analytic Network Process, the Best-Worst Method,\cite{rezaei2015bwm} fuzzy extensions) relax specific AHP assumptions but share the core logic of structured criterion weighting. This stream has not been applied to AI-integration-depth-conditioned probability estimation: existing MCDM--real-options hybrids treat technology type generically without modelling how integration depth restructures the assessment criteria themselves.

Venture-finance and private-market evidence adds a further missing element: pricing mechanics. Staged financing is modelled as milestone-contingent commitment under uncertainty,\cite{lerner2020vc,talmor2005staging} and security design drives systematic divergence between headline round valuations and common-share fair values in unicorn markets.\cite{gornall2020squaring} Regulatory and tax frameworks formalise this divergence through distinct valuation tracks for preferred and common equity.\cite{us26cfr1409a1,maplebear2023s1correspondence} In AI markets, winner-take-most dynamics and complementary-asset moats concentrate option value in a small number of firms,\cite{korinek2025concentrating,azoulay2024oldmoats,rochet2003platform} compounding the disconnect between headline figures and operating-value fundamentals.

Cross-jurisdiction evidence indicates that valuation outcomes are shaped by financing ecosystems and regulatory institutions. European competitiveness diagnostics identify persistent scale-up constraints rooted in capital-market structure, exit availability, and risk-capital depth.\cite{draghi2024competitiveness,arnold2024steppingup} EU and Swiss AI governance frameworks add a distinct layer: jurisdiction-specific compliance requirements that condition which milestones are feasible, at what cost, and on what timeline.\cite{eu2024aiact,finma2024guidance08,bakom2025aiapproach}

For the valuation framework, these requirements enter directly as $C_{\mathrm{r},s}$ in the $V_s$ decomposition and as constraints on milestone feasibility that vary by jurisdiction. A firm pursuing IDL~3 integration under EU AI Act oversight faces compliance gates that alter both the timing and the probability of milestone clearance relative to the same firm operating in a lighter regulatory environment. Regulatory jurisdiction must therefore enter the framework as a parameter that varies across cases.

No existing framework provides an industry-agnostic taxonomy that distinguishes AI integration from AI production, links both categories to milestone-gated option structures, and estimates each option's probability independently so that scenario analysis can isolate individual sensitivities. The required synthesis spans depth-conditioned classification, real-options architecture, structured per-option probability estimation, security-design effects, and jurisdiction-conditioned feasibility. Each literature stream supplies components; none combines them into a single auditable valuation logic.

\section{Contribution}

DCF remains the necessary foundation for firm valuation; the issue is which protocol governs the assumptions embedded in it when AI integration is priced. The framework developed here decomposes the opaque growth-rate assumption into identifiable options and stated probabilities, thereby making the source of the valuator's projection uncertainty ($\sigma$ in \Cref{fig:dcf-uncertainty}) explicit. The contribution is structural rather than predictive. The framework extends DCF with a decomposition protocol for AI-related inputs, without any claim to estimate option values precisely.

The scope is deliberately diagnostic and aims to make assumptions comparable across analysts, firms, and regulatory environments. The paper contributes:

\begin{enumerate}
  \item An \emph{industry-agnostic taxonomy} separating AI Integrators (IDL~0--3) from AI Providers (AI~Wrapper and AI~Native). IDL~0 and IDL~1 are valuation-neutral under this framework; IDL~2 and IDL~3 generate distinct option structures.
  \item A \emph{milestone-gated real-options overlay} where each milestone along the integration path gates a sequential investment decision with depth-specific and category-specific probabilities and payoffs, producing the additive structure $V_0 = V_{\mathrm{DCF}} + V_{\mathrm{AI}}$.
  \item A \emph{$V_s$ decomposition} splitting each milestone's opaque value into five auditable components: baseline value, operating uplift, continuation optionality, regulatory cost, and execution cost.
  \item A \emph{per-option probability architecture} where each option carries an independently estimated $p_s$ derived from AHP-based intersubjective scoring (the Success Readiness Index), calibrated against depth-specific base rates and adjustable individually in scenario analysis. The SRI protocol accommodates mixed rater panels of internal and external experts; rater pool composition is an explicit, documented design choice governed by two credentialing dimensions (domain knowledge and conflict of interest), making rater governance itself auditable and contestable.
  \item A \emph{scenario simulation layer} producing structured valuation bands where each bound traces to identifiable option-level assumptions, enabling the valuator to locate where sensitivity concentrates.
  \item \emph{Testable predictions} on when markets react to AI milestones, how analyst dispersion changes with integration depth, and which milestone types move value most.
\end{enumerate}

The framework applies at any lifecycle stage, from pre-investment planning to M\&A due diligence on firms with fully completed AI integration, where option value has migrated into the DCF baseline.

\section{Research}

The valuation problem differs structurally depending on whether a firm uses AI to transform an existing business or sells AI as a product. A bank deploying machine-learning models for credit scoring faces integration risk, organisational adaptation cost, and regulatory compliance gates; a firm that develops and licenses the underlying model faces technology risk, platform economics, and competitive moat dynamics.

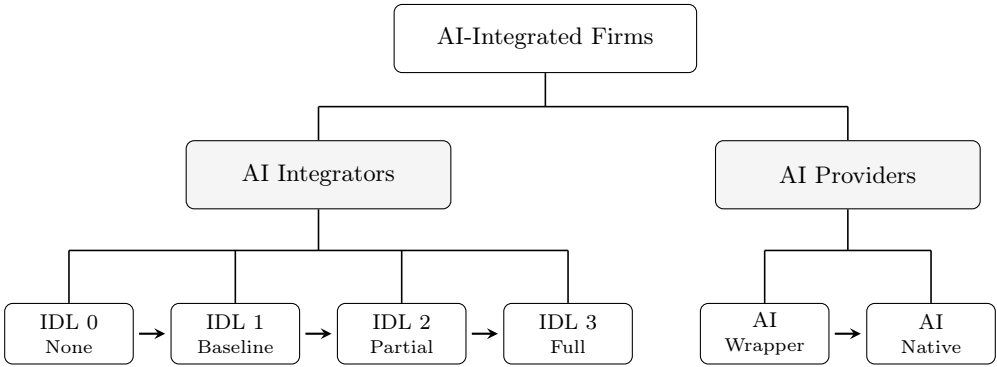
\begin{figure}[htbp]
\centering
\begin{tikzpicture}[
  rootbox/.style={draw, rounded corners=3pt, minimum height=0.9cm, minimum width=4.0cm,
                  align=center, font=\small},
  catbox/.style={draw, rounded corners=3pt, minimum height=0.9cm, minimum width=3.5cm,
                 align=center, font=\small, fill=black!4},
  leafbox/.style={draw, rounded corners=3pt, minimum height=0.8cm, minimum width=1.7cm,
                  align=center, font=\footnotesize},
  conn/.style={semithick},
  prog/.style={->, >=stealth, thick, shorten >=2pt, shorten <=2pt},
]

\node[rootbox] (root) at (0,0) {AI-Integrated Firms};

\node[catbox] (catA) at (-3.0,-1.8) {AI Integrators};
\node[catbox] (catB) at (4.0,-1.8) {AI Providers};

\coordinate (brR) at (0,-0.9);
\draw[conn] (root.south) -- (brR);
\draw[conn] (brR -| catA.north) -- (brR -| catB.north);
\draw[conn] (brR -| catA.north) -- (catA.north);
\draw[conn] (brR -| catB.north) -- (catB.north);

\node[leafbox] (idl1) at (-6.3,-3.9) {IDL~0\\[-0.1em]{\scriptsize None}};
\node[leafbox] (idl2) at (-4.1,-3.9) {IDL~1\\[-0.1em]{\scriptsize Baseline}};
\node[leafbox] (idl3) at (-1.9,-3.9) {IDL~2\\[-0.1em]{\scriptsize Partial}};
\node[leafbox] (idl4) at (0.3,-3.9)  {IDL~3\\[-0.1em]{\scriptsize Full}};

\coordinate (brA) at (-3.0,-2.8);
\draw[conn] (catA.south) -- (brA);
\draw[conn] (brA -| idl1.north) -- (brA -| idl4.north);
\draw[conn] (brA -| idl1.north) -- (idl1.north);
\draw[conn] (brA -| idl2.north) -- (idl2.north);
\draw[conn] (brA -| idl3.north) -- (idl3.north);
\draw[conn] (brA -| idl4.north) -- (idl4.north);

\draw[prog] (idl1) -- (idl2);
\draw[prog] (idl2) -- (idl3);
\draw[prog] (idl3) -- (idl4);

\node[leafbox] (typeW) at (2.9,-3.9) {AI\\[-0.1em]{\scriptsize Wrapper}};
\node[leafbox] (typeN) at (5.1,-3.9) {AI\\[-0.1em]{\scriptsize Native}};

\draw[prog] (typeW) -- (typeN);

\coordinate (brB) at (4.0,-2.8);
\draw[conn] (catB.south) -- (brB);
\draw[conn] (brB -| typeW.north) -- (brB -| typeN.north);
\draw[conn] (brB -| typeW.north) -- (typeW.north);
\draw[conn] (brB -| typeN.north) -- (typeN.north);

\end{tikzpicture}
\caption{Taxonomy of AI-integrated firms: AI Integrators by depth (IDL~0--3) and AI Providers by type, with progression arrows.}
\label{fig:taxonomy}
\end{figure}

A single firm may contain elements of both, but the valuation logic for each component differs. Collapsing both into one classification obscures the mechanisms that drive option value. \Cref{fig:taxonomy} introduces a taxonomy that separates these categories and defines the sub-classifications used throughout this paper.

AI Integrators are classified by Integration Depth Level (IDL~0--3). IDL~0 denotes no AI involvement and serves as a boundary condition. IDL~1 (Baseline) captures firms using off-the-shelf AI tools such as general-purpose chatbots or embedded copilot features in standard software; this creates no switching costs, no proprietary capability, and no measurable cash-flow restructuring. Because neither IDL~0 nor IDL~1 generates a decision node with identifiable conditional value, the option overlay is zero by construction and the framework's analytical contribution begins at IDL~2. IDL~2 (Partial) marks the threshold where AI shapes specific business processes, producing measurable efficiency gains and the first real options. IDL~3 (Full) describes firms where AI is embedded across core operations, trained on proprietary data, and generates strategic feedback loops. At this depth, continuation-option value dominates operating gains, and the valuation problem shifts from measuring efficiency improvements to pricing future strategic optionality. The classification is cross-sectional; firms at any IDL level may remain there indefinitely.

AI Providers are subdivided into AI~Wrapper and AI~Native. A wrapper builds on third-party models, adding domain-specific interfaces or vertical integration without developing the underlying technology. A native develops proprietary models or infrastructure as its core product. The risk profiles diverge: wrappers face upstream dependency and compete on application-layer differentiation; natives face deep technology risk but can build platform-level moats. Progression from wrapper to native is rare but real, and the distinction matters for valuation. A permanent wrapper carries a narrower option set. A wrapper with a credible plan to develop proprietary models carries substantially more option value, because the transition unlocks moats, pricing power, and reduced dependency on upstream providers.

The unit of analysis for the valuation model is the \emph{milestone state}: an observable, dateable point in the AI integration path at which the firm faces a decision node (continue, expand, pivot, or abandon). Each milestone state $s$ carries a conditional value $V_s$ and an independently estimated probability of success $p_s$; the model prices the firm's AI component as a portfolio of such milestone-gated options. \Cref{tab:idl-overview} summarises the valuation implications by integration depth.

{\footnotesize
\begin{xltabular}{\textwidth}{l X X l l}
\caption{Valuation implications by integration depth level.}
\label{tab:idl-overview} \\
\toprule
IDL & DCF impact & Option relevance & Risk & Valuator focus \\
\midrule
\endfirsthead
\multicolumn{5}{l}{\footnotesize\tablename~\thetable{} -- continued} \\
\toprule
IDL & DCF impact & Option relevance & Risk & Valuator focus \\
\midrule
\endhead
\bottomrule
\endfoot
0 (None) & Negligible & None & None & Ignore \\
1 (Baseline) & Negligible & None & None & Monitor \\
2 (Partial) & Margin improvement & Expand/abandon & Moderate & Core driver \\
3 (Full) & CF restructuring & Build/license/spin-off & High & Entire valuation \\
\end{xltabular}
}

The framework proposed here builds on DCF and makes the assumptions an analyst already embeds in it explicit and decomposable. In standard practice, the AI-related growth premium is a single opaque parameter. The option overlay replaces that parameter with a structured set of milestone-level inputs: which milestones are priced, at what probability, with what conditional value, and under which IDL classification. A sufficiently rich multi-scenario DCF with probability-weighted stage gates approximates this structure; at that point, the analyst is performing real-options analysis with DCF mechanics. The contribution is a transparent architecture that makes the assumptions behind the number auditable and comparable across analysts.

\subsection{Valuation model}
\label{sec:milestone-model}

The total firm value separates existing operations from AI-related option value:
\begin{equation}\label{eq:v-total}
V_0 \;=\; V_{\mathrm{DCF}} \;+\; V_{\mathrm{AI}}
\end{equation}
where $V_{\mathrm{DCF}}$ is the discounted cash-flow value of the firm's existing operations and $V_{\mathrm{AI}}$ is the real-options overlay capturing the value of staged AI integration decisions. The AI component aggregates across all milestone-gated options:
\begin{equation}\label{eq:v-ai}
V_{\mathrm{AI}} \;=\; \sum_{s=1}^{n} \frac{p_s \cdot V_s}{(1+r_s)^{T_s}} \;-\; I_0
\end{equation}
where $n$ is the number of identified options, $p_s$ the probability of reaching state $s$, $V_s$ the value conditional on reaching it, $T_s$ the expected time to reach $s$, $r_s$ the risk-adjusted discount rate for option $s$, and $I_0$ the initial investment already committed.

\Cref{eq:v-total,eq:v-ai} decompose the opaque growth-rate assumption into explicit components. The parameterisation is more demanding, but it forces the valuator to state which milestones are priced, at what probability, and with what conditional value.

The value conditional on reaching milestone state $s$ decomposes into five components:
\begin{equation}\label{eq:vs-decomp}
V_s \;=\; V_{\mathrm{b},s} \;+\; \Delta V_{\mathrm{o},s} \;+\; V_{\mathrm{c},s} \;-\; C_{\mathrm{r},s} \;-\; C_{\mathrm{e},s}
\end{equation}
where $V_{\mathrm{b},s}$ is the baseline business value given sunk costs, $\Delta V_{\mathrm{o},s}$ the incremental operating value from measurable cash-flow changes, $V_{\mathrm{c},s}$ the continuation-option value from future decisions that only exist because state $s$ was reached, $C_{\mathrm{r},s}$ the regulatory and compliance cost, and $C_{\mathrm{e},s}$ the execution cost including integration debt, retraining, and organisational friction.

The five-component decomposition makes the structural shift across integration depths visible. At IDL~2, measurable operating gains ($\Delta V_{\mathrm{o},s}$) dominate: a workflow redesign produces quantifiable cost savings or revenue gains. At IDL~3, continuation-option value ($V_{\mathrm{c},s}$) dominates: the firm's value depends on future decisions about platform scaling, new verticals, and data network effects that have not yet been made. \Cref{tab:vs-ratio} summarises this shift. The implication for valuation practice is direct: analyst dispersion increases with integration depth because $V_{\mathrm{c},s}$ is inherently harder to estimate than $\Delta V_{\mathrm{o},s}$.

{\footnotesize
\begin{xltabular}{\textwidth}{l X X X}
\caption{$V_s$ composition by integration depth.}
\label{tab:vs-ratio} \\
\toprule
 & IDL~2 & IDL~3 & AI Provider \\
\midrule
\endfirsthead
\multicolumn{4}{l}{\footnotesize\tablename~\thetable{} -- continued} \\
\toprule
 & IDL~2 & IDL~3 & AI Provider \\
\midrule
\endhead
\bottomrule
\endfoot
$\Delta V_{\mathrm{o},s}$ weight & Dominant & Moderate & Varies \\
$V_{\mathrm{c},s}$ weight & Large & Dominant & Dominant \\
Estimation difficulty & Medium--high & High & High \\
Analyst dispersion & High & Very high & Very high \\
\end{xltabular}
}

Estimation of $V_s$ components differs by integration depth. At IDL~2, pilot data and operational metrics provide direct estimates of $\Delta V_{\mathrm{o},s}$: measured cost reductions in one department extrapolate to comparable departments through standard capital-budgeting methods. Typical milestones at this depth (proof of concept, pilot deployment, workflow integration, expansion decision) are shorter, less costly, and carry lower payoff variance. Continuation-option value remains small, valued as $p_{\mathrm{expand}} \times \Delta V_{\mathrm{next}}$. At IDL~3, direct measurement gives way to scenario modelling. Milestones (data infrastructure build-out, model training, production deployment, competitive differentiation) take longer, cost more, and carry higher payoff variance. $V_{\mathrm{c},s}$ involves strategic optionality (license the model, spin off the AI unit, enter adjacent markets), each priced as a separate mini-option anchored against comparable transactions. For AI providers, the biotech pipeline analogy applies: phase-gated investment with binary milestone outcomes at each gate, estimated from revenue traction, addressable market models, and funding-round repricing events.

A firm operating at different IDL levels across business functions (a bank at IDL~2 in credit scoring, IDL~0 in back-office correspondence) is treated as a portfolio: $V_{\mathrm{AI}} = \sum_{f} V_{\mathrm{AI},f}$, where $f$ indexes business functions and each function is classified and valued independently. Functions at IDL~0 and IDL~1 contribute zero to the option overlay.

\subsection{Probability architecture}
\label{sec:probability}

The probability $p_s$ in \eqref{eq:v-ai} is the most vulnerable parameter in any real-options application; in practice, it is often an unstructured expert guess. The protocol proposed here replaces that guess with an auditable, intersubjective estimate anchored in a multi-criteria framework. Its core instrument is the \emph{Success Readiness Index} (SRI): a normalised score derived from Analytic Hierarchy Process (AHP) based expert assessment, reflecting how ready an option is to succeed. The SRI is a priority score on a ratio scale measuring relative readiness across criteria, not calibrated probability. Multiple experts score each option independently via the same AHP hierarchy; results are aggregated using the geometric mean, preserving the ratio scale per Saaty's group-aggregation protocol. The per-option probability $p_s$ is then obtained by calibrating the SRI against IDL-specific base rates derived from comparable technology adoption outcomes. Each option carries its own $p_s$, enabling independent scenario variation across the portfolio: the SRI provides the ordinal ranking; the calibration step maps it onto a probability scale.

The rater panel may include both internal experts (founders, technical leads, domain owners) and external experts (independent advisers, sector specialists). Insiders contribute system-specific knowledge on data infrastructure, organisational constraints, and technical feasibility that external raters would struggle to replicate in a short engagement. The relevant distinction lies in two independent dimensions of each rater's contribution: domain knowledge $w_i^{\mathrm{comp}} \in [0,1]$ (expertise on this specific option type) and conflict of interest $b_i \in [0,1]$ (financial stake, reputational commitment, or prior public positions on the firm). These yield an effective aggregation weight
\begin{equation}\label{eq:w-eff}
w_i^{\mathrm{eff}} \propto w_i^{\mathrm{comp}} \cdot (1 - \lambda b_i),
\end{equation}
where $\lambda \in [0,1]$ governs how aggressively the protocol penalizes conflicts of interest. Both scores and the resulting weights are documented alongside the SRI output, so a reviewer can challenge any weight assignment, substitute alternatives, and trace the effect on the aggregated score. Rater pool composition becomes an auditable design choice rather than an implicit assumption; \Cref{tab:rater-rubric} provides the scoring rubric for both dimensions.

{\footnotesize
\begin{xltabular}{\textwidth}{l X X}
\caption{Rater credentialing rubric: $w_i^{\mathrm{comp}}$ (domain knowledge) and $b_i$ (conflict of interest).}
\label{tab:rater-rubric} \\
\toprule
Band & $w_i^{\mathrm{comp}}$; domain knowledge & $b_i$; conflict of interest \\
\midrule
\endfirsthead
\multicolumn{3}{l}{\footnotesize\tablename~\thetable{} -- continued} \\
\toprule
Band & $w_i^{\mathrm{comp}}$; domain knowledge & $b_i$; conflict of interest \\
\midrule
\endhead
\bottomrule
\endfoot
0.8--1.0 & Deep operational expertise in this specific option type; multiple comparable prior engagements; demonstrable track record & Direct financial stake (equity, options, success fee); public prior endorsement of the valuation outcome; lead investor or co-founder \\
0.5--0.7 & Solid sector knowledge; some comparable experience; relevant professional credentials & Indirect financial interest (adviser fee, ongoing contract with firm); prior public statements favouring a specific outcome \\
0.2--0.4 & General industry knowledge; limited direct experience with this option type; familiarity from adjacent domains & Reputational interest only; no financial stake; known association with firm or management \\
0.0--0.1 & No relevant domain expertise; generalist background only & No financial stake; no prior public positions on this firm or option; no ongoing relationship \\
\end{xltabular}
}

The protocol proceeds in three stages. In the first, experts score each option against five criteria through AHP pairwise comparisons: technical feasibility, regulatory compliance, organisational readiness, market impact potential, and data access and governance. The output is a weighted priority vector constituting the SRI. Consistency ratios flag incoherent judgements, and multi-rater aggregation via geometric mean surfaces disagreements invisible in single-rater estimates.

In the second stage, the SRI is calibrated against IDL-specific base rates to produce $p_s$. A high SRI does not automatically imply a high $p_s$: a deep-integration option at IDL~3 may score well on readiness criteria but face a structurally lower base rate because the underlying milestone is more complex and less precedented. The calibration anchors are derived from comparable technology adoption rates, M\&A stage conversion rates, or venture-capital milestone outcomes. These anchors are illustrative priors, not empirically derived base rates; they set the probability scale. In the third stage, scenario analysis shifts $p_s$ values (selectively, across the board, or in targeted combinations) to produce a valuation band. The second stage sets the scale; the third stage stress-tests it.

The AHP criterion weights are not fixed across integration depths. At IDL~2, technical feasibility and organisational readiness carry the largest weight: the central question is whether the firm can build and absorb the AI system. At IDL~3 and for AI providers, market impact potential and regulatory risk dominate: the technology is assumed capable, and value depends on market acceptance and regulatory permission. This depth-dependent weight shift reflects the structural reality that the binding constraint on option exercise changes with integration depth.

The consistency ratio deserves explicit framing. The CR measures internal judgement coherence across pairwise comparisons; it does not measure empirical truth. A low CR (below the conventional 0.10 threshold) indicates that the rater's comparative judgements are logically consistent; a high CR flags structural incoherence requiring re-elicitation. The hypothesis that higher-IDL options produce greater inter-rater dispersion in SRI scores, together with higher average CRs, is testable and follows from the structural prediction that $V_{\mathrm{c},s}$-dominated options are harder to assess. This paper frames the hypothesis as exploratory; confirming or rejecting it requires the empirical data collection described in \cref{sec:empirical}.

Because each option carries its own $p_s$, the framework supports targeted scenario analysis. An analyst can shift the probability of a single option, a thematic group, or the entire portfolio and trace the effect through to $V_{\mathrm{AI}}$. Selective shifts isolate sensitivity to specific assumptions (a regulatory tightening, a technology breakthrough); blanket shifts produce ceiling and floor estimates; mixed scenarios combine both. The output is a valuation band whose bounds trace to identifiable option-level assumptions. If shifting one option's $p_s$ from 0.3 to 0.6 moves $V_{\mathrm{AI}}$ by 40\%, that option is where risk concentrates.

When a firm pursues multiple AI options, their effects may overlap. If Option~A (AI customer scoring) and Option~B (AI pricing) both claim credit for the same revenue uplift, summing their $V_s$ values double-counts the gain. The framework must address this without collapsing into a combinatorial problem that defeats practical application.

Three approaches exist, each with a different precision--practicality trade-off (\cref{tab:interdependency}). Attribution by assignment requires the valuator to allocate each value effect to one option, or to split it explicitly with documented percentages. The interaction matrix is an $n \times n$ diagnostic that flags option pairs as synergistic, overlapping, or independent, forcing the valuator to consider interactions before summing. Shapley value decomposition computes each option's average marginal contribution across all possible coalitions; it is formally correct but computationally prohibitive for practical portfolios ($n = 10$ generates 1\,024 coalition evaluations).

{\footnotesize
\begin{xltabular}{\textwidth}{l X l l}
\caption{Approaches to option interdependency by mechanism, practicality, and precision.}
\label{tab:interdependency} \\
\toprule
Method & Mechanism & Practicality & Precision \\
\midrule
\endfirsthead
\multicolumn{4}{l}{\footnotesize\tablename~\thetable{} -- continued} \\
\toprule
Method & Mechanism & Practicality & Precision \\
\midrule
\endhead
\bottomrule
\endfoot
Attribution by assignment & Human allocation of value effects & High & Medium \\
Interaction matrix & $n \times n$ synergy/overlap flags & High & Low \\
Shapley values & Marginal contribution across coalitions & Low & High \\
\end{xltabular}
}

The framework proposes attribution by assignment as the core rule, supplemented by the interaction matrix as a diagnostic check. Shapley values remain a direction for future work.

\subsection{Lifecycle and empirical design}
\label{sec:lifecycle}

The framework applies at any point in the AI integration lifecycle, including post-milestone reassessment. As milestones are cleared, option value migrates into the DCF baseline:
\begin{equation}\label{eq:v-lifecycle}
V_{\mathrm{firm},t} \;=\; V_{\mathrm{DCF},t} \;+\; V_{\mathrm{AI},t}
\end{equation}
where $V_{\mathrm{DCF},t}$ absorbs the cash-flow contributions of exercised options and $V_{\mathrm{AI},t}$ contracts as uncertainty resolves. At any valuation date, each option falls into one of four status categories, summarised in \cref{tab:option-status}.

{\footnotesize
\begin{xltabular}{\textwidth}{l X X}
\caption{Option status classification at a given valuation date.}
\label{tab:option-status} \\
\toprule
Status & Value location & Treatment \\
\midrule
\endfirsthead
\multicolumn{3}{l}{\footnotesize\tablename~\thetable{} -- continued} \\
\toprule
Status & Value location & Treatment \\
\midrule
\endhead
\bottomrule
\endfoot
Not started & Fully in $V_{\mathrm{AI}}$ & Full AHP $\to$ SRI $\to$ $p_s$ $\to$ scenario band \\
In progress & Split: cleared in DCF, remaining in options & Reduced $V_{\mathrm{AI}}$; $p_s$ updated \\
Completed & Fully in $V_{\mathrm{DCF}}$ & Measurable cash-flow contribution \\
New continuation & In $V_{\mathrm{AI}}$ (new) & Fresh AHP scoring \\
\end{xltabular}
}

Probabilities are not static. As milestones clear, $p_s$ for the next milestone updates upward (demonstrated capability reduces uncertainty), $r_s$ decreases (risk resolved at each gate), and new continuation options may emerge that were not identifiable at the earlier valuation date. A firm that has cleared four of five milestones carries a structurally higher $p_5$ than one that has not started. This Bayesian updating is the mechanism behind the endogenous risk shift: clearing milestones resolves risk and reprices the remaining option portfolio. The framework treats this updating as conceptual; specifying a functional form for $p_{s+1 \mid s\;\text{cleared}}$ would risk overspecification for a framework paper and is deferred to future work.

In an acquisition context, buyer and seller see different option sets. The seller's $V_{\mathrm{AI}}$ reflects options exercisable with the seller's current resources, organization, and market position. The buyer's $V_{\mathrm{AI}}$ reflects options enabled by the combination: cross-selling into the buyer's customer base, scaling AI to the buyer's operations, or leveraging the buyer's data assets. The synergy premium becomes structurally transparent:
\begin{equation}\label{eq:synergy}
\Delta V_{\mathrm{synergy}} \;=\; V_{\mathrm{AI}}^{\mathrm{buyer}} \;-\; V_{\mathrm{AI}}^{\mathrm{seller}}
\end{equation}
This decomposition reduces M\&A negotiation from a single headline disagreement to a structured comparison of specific options, probabilities, and continuation values that each party prices differently.

\subsection{Illustrative application}
\label{sec:empirical}

The framework generates testable predictions: (a)~milestone type and IDL depth predict the magnitude of market reactions to AI integration announcements, (b)~higher-IDL milestones produce larger abnormal returns, (c)~later-stage milestones within a given IDL level produce diminishing incremental effects as option value is progressively priced in, and (d)~valuation dispersion increases monotonically with integration depth.

The framework is illustrated through a single evidence-anchored case reconstruction, parameterized from a real firm. The goal is to show that the model produces coherent, decomposable valuation bands under transparent, citable inputs, not to claim causal validation against market prices. All inputs are stated explicitly so readers can verify the arithmetic and substitute their own assumptions.
\label{sec:case-study}

\emph{Case Firm E12} is a Swiss AI-native SaaS startup building an automated energy portfolio optimisation platform.\footnote{The firm's identity is anonymised in accordance with data-protection requirements and standard research-ethics guidelines for confidential primary-source case studies. The underlying primary documents (investor masterplan, financial model, milestone register, TRL roadmap) are available from the authors upon reasonable request, subject to a non-disclosure agreement.} Its core product, E12-eAI, trades simultaneously across Day-Ahead, Intraday, Futures, power purchase agreements, flexibility markets, and CO\textsubscript{2} certificates, replacing the manual spreadsheets and fragmented legacy tools that most energy portfolio operators still rely on. Revenues combine platform licences, performance fees tied to realised customer margin uplift, data products, and integration services. Under the framework taxonomy, Case Firm E12 is an \emph{AI Provider (Native)}: E12-eAI is the product sold to customers, not a tool deployed internally; the energy portfolio operators who adopt the platform are the AI Integrators.

\begin{figure}[H]
\centering
\begin{tikzpicture}[>=stealth, font=\small, x=1.2cm]

\draw[->, thick] (-0.2, 0) -- (10.9, 0);

\fill[gray!20] (1.25,0) circle (0.38cm); \draw[thick]      (1.25,0) circle (0.38cm);
\node[font=\footnotesize] at (1.25,0) {M1};

\fill[gray!60] (2.5,0)  circle (0.38cm); \draw[thick]      (2.5,0)  circle (0.38cm);
\node[font=\footnotesize, text=white] at (2.5,0) {M2};

\fill[gray!20] (5.0,0)  circle (0.38cm); \draw[thick]      (5.0,0)  circle (0.38cm);
\node[font=\footnotesize] at (5.0,0) {M3};

\fill[gray!40] (10.0,0) circle (0.38cm); \draw[thick]      (10.0,0) circle (0.38cm);
\node[font=\footnotesize] at (10.0,0) {M4};

\node[text width=2.8cm, align=center, font=\footnotesize] at (2.5, 1.1)
  {2~paid pilots\\CHF~13.5m tranche\\[1pt]{\itshape\color{gray!65}financing gate}};

\node[text width=2.8cm, align=center, font=\footnotesize] at (10.0, 1.1)
  {10+~customers\\TRL~7 model\\reprod.\ GTM};

\node[text width=2.5cm, align=center, font=\footnotesize] at (1.25, -1.1)
  {Shadow mode\\2~API pilots};

\node[text width=2.5cm, align=center, font=\footnotesize] at (5.0, -1.1)
  {CHF~400k ARR\\first revenues};

\foreach \x/\mo in {0/0, 1.25/3, 2.5/6, 5.0/12, 10.0/24} {
  \draw[gray!45, thin] (\x, -1.7) -- (\x, -1.85);
  \node[font=\footnotesize] at (\x, -2.05) {\mo};
}
\node[right, font=\footnotesize, text=gray!60] at (10.3, -1.85) {Month};

\draw[gray!45, semithick] (0, -2.6) -- (10.0, -2.6);
\foreach \x/\trl in {0/2--3, 1.25/4, 2.5/5, 5.0/6+, 10.0/7} {
  \fill[gray!45] (\x, -2.6) circle (2pt);
  \node[below=2pt, font=\footnotesize, text=gray!60] at (\x, -2.62) {TRL~\trl};
}
\end{tikzpicture}
\caption{Milestone gate structure and TRL progression, Case Firm E12 (Q2~2026--Q2~2028). Source: \autocite{firmE12_masterplan}.}
\label{fig:firmE12-timeline}
\end{figure}
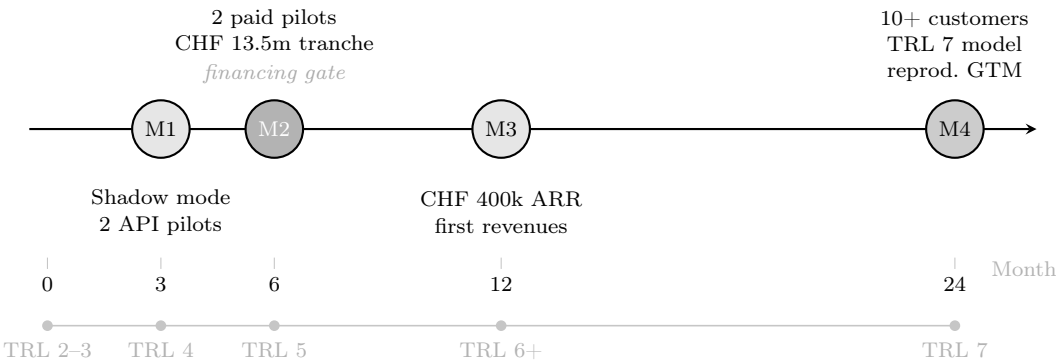

At the valuation date (Q2 2026), the AI core is at TRL~2--3: the optimisation algorithm has been validated on real market data, but the agent-orchestration layer is still in pre-development. The firm is targeting a CHF~15 million Seed round at a pre-money valuation of CHF~100 million. Four milestone gates over 24 months structure both the technical roadmap and investor decision rights (\cref{fig:firmE12-timeline}): Gate~1 (month~3) confirms system stability for customer-facing pilots; Gate~2 (month~6) is the financing trigger, releasing the remaining CHF~13.5 million tranche only if two pilots have converted to paid agreements; Gate~3 (month~12) validates product-market fit through first recurring revenues (CHF~400k ARR); Gate~4 (month~24) establishes a reproducible go-to-market model with a proprietary TRL~7 model in production and 10+ active customers. Gate failure closes capital access and terminates all downstream options.

Each gate maps directly onto the option structure in \cref{eq:v-ai}: milestone state $s$, conditional value $V_s$, and sequential dependency such that failure at any gate terminates downstream options. The $V_s$ decomposition (\cref{eq:vs-decomp}) applies at each gate. In \cref{tab:firmE12}, $w_s = p_s V_s / (1+r_s)^{T_s}$ denotes the present-value-weighted contribution of each option to $V_{\mathrm{AI}}$, and $|\delta w_s|$ captures the absolute change in $w_s$ from a uniform $\pm 0.10$ shift in $p_s$, holding other milestones at base. $V_s$ combines all five components of \cref{eq:vs-decomp}; $\Delta V_{\mathrm{o},s}$ and $V_{\mathrm{c},s}$ are shown separately as the dominant drivers. All values are in CHF millions.

{\footnotesize
\begin{xltabular}{\textwidth}{X r r r r r r r r}
\caption{Milestone-gated valuation reconstruction, Case Firm E12 (Q2 2026). Source: \autocite{firmE12_masterplan}.}
\label{tab:firmE12} \\
\toprule
Milestone & $T_s$ & $p_s$ & $\Delta V_{\mathrm{o},s}$ & $V_{\mathrm{c},s}$ & $V_s$ & $r_s$ & $w_s$ & $|\delta w_s|$ \\
& (yr) & & (CHF m) & (CHF m) & (CHF m) & & (CHF m) & (CHF m) \\
\midrule
\endfirsthead
\multicolumn{9}{l}{\tablename~\thetable{} -- continued} \\
\toprule
Milestone & $T_s$ & $p_s$ & $\Delta V_{\mathrm{o},s}$ & $V_{\mathrm{c},s}$ & $V_s$ & $r_s$ & $w_s$ & $|\delta w_s|$ \\
& (yr) & & (CHF m) & (CHF m) & (CHF m) & & (CHF m) & (CHF m) \\
\midrule
\endhead
\bottomrule
\endfoot
M1: Shadow mode stable; 2 API pilots in onboarding (TRL~4) & 0.25 & 0.82 & 2.0 & 5.0 & 13.9 & 15\% & 11.0 & 1.3 \\
M2: 2 active paid pilots; CHF 13.5m tranche unlocked (TRL~5) & 0.50 & 0.67 & 6.0 & 15.0 & 34.9 & 20\% & 21.4 & 3.2 \\
M3: First recurring revenues (CHF~400k ARR); TRL~6+ & 1.00 & 0.52 & 18.0 & 35.0 & 79.3 & 25\% & 33.0 & 6.3 \\
M4: Reproducible GTM; 10$+$ customers; TRL~7 model & 2.00 & 0.38 & 48.0 & 95.0 & 198.5 & 30\% & 44.6 & 11.7 \\
\midrule
$\sum w_s$ & & & & & & & 110.0 & \\
$-I_0$ (seed capital committed) & & & & & & & $-$15.0 & \\
$V_{\mathrm{AI}}$ & & & & & & & \textbf{95.0} & \\
$V_{\mathrm{DCF}}$ (IP/team baseline) & & & & & & & 5.0 & \\
$V_0 = V_{\mathrm{DCF}} + V_{\mathrm{AI}}$ & & & & & & & \textbf{100.0} & \\
\end{xltabular}
}

\begin{figure}[htbp]
\centering
\begin{tikzpicture}
\begin{axis}[
  xbar,
  width=0.68\textwidth,
  height=5.2cm,
  bar width=0.55cm,
  symbolic y coords={M4,M3,M2,M1},
  ytick=data,
  yticklabel style={font=\small},
  xlabel={$w_s$, PV-weighted option contribution (CHF~m)},
  xlabel style={font=\small},
  ticklabel style={font=\small},
  xmin=0, xmax=70,
  enlarge y limits=0.3,
]
\addplot+[
  fill=gray!35, draw=gray!60,
  error bars/.cd,
  x dir=both, x explicit,
  error bar style={line width=1.2pt, gray!60},
  error mark options={rotate=90, mark size=5pt, line width=1.2pt, gray!60},
] coordinates {
  (44.6,M4) +- (11.7,0)
  (33.0,M3) +- (6.3,0)
  (21.4,M2) +- (3.2,0)
  (11.0,M1) +- (1.3,0)
};
\node[anchor=west, font=\footnotesize] at (axis cs:58.0,M4) {44.6};
\node[anchor=west, font=\footnotesize] at (axis cs:41.0,M3) {33.0};
\node[anchor=west, font=\footnotesize] at (axis cs:26.0,M2) {21.4};
\node[anchor=west, font=\footnotesize] at (axis cs:14.0,M1) {11.0};
\end{axis}
\end{tikzpicture}
\caption{PV-weighted option contributions $w_s$ per milestone with $\pm 0.10$ sensitivity bands ($|\delta w_s|$).}
\label{fig:firmE12-sensitivity}
\end{figure}
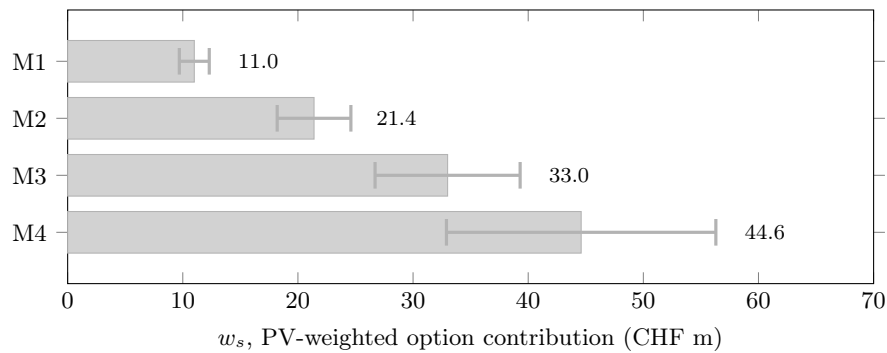

Case Firm E12 has no revenue history at the valuation date. That means $V_{\mathrm{b},s}$, the baseline value given sunk costs, is carried by team credentials and IP rather than by cashflow extrapolation, which is why the $\Delta V_{\mathrm{o},s}$ values at M1 and M2 are small: no sustained customer margin uplift has been demonstrated yet. Value is almost entirely forward-looking, concentrated in $V_{\mathrm{c},s}$, the competitive-moat option that unlocks once proprietary TRL~7 models are in production and deliver continuously improving performance advantages over competitors still dependent on external-model wrappers. The option-killing gate structure has real teeth: the CHF~13.5 million tranche is contractually tied to Gate~2 pilot conversion, so two missed pilots terminate most available capital and effectively kill M3 and M4. The AHP--SRI probability priors reflect this structure directly: $p_{\mathrm{M1}} = 0.82$ captures near-complete technical control over Gate~1 conditions; $p_{\mathrm{M4}} = 0.38$ captures the compounded uncertainty of product-market fit, competitive dynamics, and 24 months of model maturation.

The base case yields CHF~100 million, matching the disclosed pre-money exactly. The inputs were derived independently from the masterplan's financial model and gate definitions, so the convergence reflects genuine internal consistency. For context, the masterplan itself arrives at the same figure through a distinct methodology: a milestone-scoring overlay across 13 strategic components (readiness score 36.9\% in Q2~2026; strategic value CHF~102.1m) is added to a negative DCF contribution (WACC~30\%, pre-revenue; $-$CHF~1.3m), yielding CHF~100.8m, rounded to CHF~100m as the pre-money anchor (\autocite{firmE12_masterplan}, Ch.~8). Two independent methods converging on the same value provides stronger evidence of internal consistency than either alone. The full valuation band runs from CHF~77 million (downside: all $p_s - 0.10$) to CHF~123 million (upside: all $p_s + 0.10$), a range of CHF~46 million. As \cref{fig:firmE12-sensitivity} shows, M4 drives more than 60\% of that range ($|\delta w_s| = 11.7$): the proprietary model maturation and reproducible GTM option concentrate risk in the later stage, exactly as the structural prediction for AI Providers anticipates. Removing M4 entirely collapses $V_{\mathrm{AI}}$ from CHF~95 million to approximately CHF~50 million, meaning that roughly half of firm value rests on 24-month execution. Two analysts who disagree on the CHF~100 million pre-money can now identify whether the disagreement sits in $p_{\mathrm{M4}}$, in $V_{\mathrm{c},\mathrm{M4}}$, or in the 30\% discount rate, each of which is individually challengeable against observable proxies.

\section{Discussion}

Standard valuation approaches do not ignore AI integration; they absorb it into an undifferentiated growth-rate assumption. The five-component $V_s$ decomposition (\cref{eq:vs-decomp}) separates option value at each integration stage. $\Delta V_{\mathrm{o},s}$ captures the structural efficiency gain realised at the IDL~2 transition, $V_{\mathrm{c},s}$ captures the competitive-moat value that becomes accessible at IDL~3, and $C_{\mathrm{r},s}$ and $C_{\mathrm{e},s}$ record compliance and execution costs instead of netting them silently against projected revenues. The contribution does not lie in a different total valuation, but in the decomposition itself. Disagreement over $V_{\mathrm{AI}}$ can then be located in a specific component, probability estimate, input value, or discount rate.

This decomposition points to two systematic mispricing channels under conventional approaches. IDL~1--2 firms are likely to be undervalued when option value is ignored: the operational efficiency gains captured by $\Delta V_{\mathrm{o},s}$ are real and staged, but a single-stage DCF with a generic AI premium does not separate them from baseline revenues. IDL~3 firms face the opposite risk: markets may price network effects and platform control, represented here by $V_{\mathrm{c},s}$, as already realised although they remain contingent on milestone completion. The framework does not remove that contingency, but it does isolate the probability assumptions ($p_s$) that drive the valuation instead of embedding them in a single discount rate.

\begin{figure}[htbp]
\centering
\begin{tikzpicture}
\begin{axis}[
  ybar stacked,
  bar width=1.8cm,
  width=0.60\textwidth,
  height=5.8cm,
  symbolic x coords={HGB,IFRS,{US GAAP}},
  xtick=data,
  x tick label style={font=\small},
  ymin=0, ymax=12,
  ytick={0,2,4,6,8,10},
  ylabel={CHF m (illustrative)},
  ylabel style={font=\small},
  ticklabel style={font=\small},
  enlarge x limits=0.35,
  legend style={
    at={(0.5,-0.22)}, anchor=north,
    legend columns=2, font=\small, draw=none,
    /tikz/every even column/.append style={column sep=0.3cm},
  },
]
\addplot+[fill=gray!60, draw=gray!75,
  nodes near coords, every node near coord/.style={font=\footnotesize,black}]
  coordinates {(HGB,10) (IFRS,4) ({US GAAP},5)};
\addplot+[fill=gray!15, draw=gray!40,
  nodes near coords, point meta=explicit symbolic,
  every node near coord/.style={font=\footnotesize,black}]
  coordinates {(HGB,0) [] (IFRS,6) [6] ({US GAAP},5) [5]};
\legend{Expensed (P\&L), Capitalised (balance sheet)}
\end{axis}
\end{tikzpicture}
\caption{Illustrative treatment of a CHF~10 million AI development investment under three reporting regimes.}
\label{fig:accounting-treatment}
\end{figure}
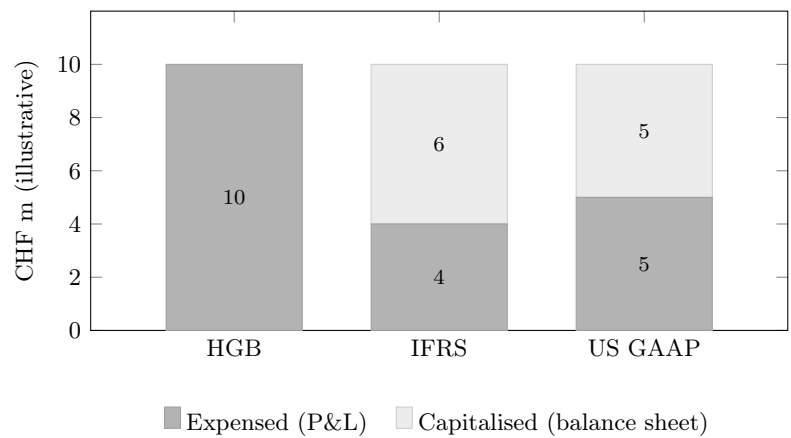

These mispricing channels are compounded by financial reporting standards. Under HGB (§248~II), development costs carry an optional capitalisation right (Aktivierungswahlrecht) that conservative preparers typically do not exercise; AI development expenditure hits the income statement in full and the balance sheet carries nothing. Under IFRS (IAS~38.57), development costs must be capitalised once six criteria are met: (i)~technical feasibility of completing the asset, (ii)~intention to complete and use or sell it, (iii)~ability to use or sell it, (iv)~demonstration of probable future economic benefits, (v)~availability of adequate technical and financial resources, and (vi)~ability to measure reliably the attributable expenditure; research costs are always expensed. The capitalisation boundary is judgement-intensive and varies across firms and auditors. Under US~GAAP (ASC~350-40 for internal-use software), capitalisation begins only at the application development stage and stops at post-implementation, with preliminary project work always expensed. The result: two firms committing CHF~10 million to economically identical AI development programmes may report dramatically different earnings, asset values, and return metrics depending solely on their reporting regime (\cref{fig:accounting-treatment}). A value driver that generates CHF~95 million in real-option value may simultaneously produce a multi-year operating loss under HGB and a partially capitalised intangible asset under IFRS. The economic reality is identical; the reported financial statements are not.

The framework addresses this directly. The cost term $C_{\mathrm{e},s}$ in the $V_s$ decomposition (\cref{eq:vs-decomp}) records AI development expenditure as an economic cash outflow, regardless of whether the applicable standard classifies it as an asset or an expense. Two analysts applying the framework to the same firm under different reporting regimes will produce the same option-level inputs, because the decomposition operates on the economic layer beneath the reporting layer. This is precisely why it produces comparable valuations where unadjusted multiple-based methods produce incomparable ones.

The structural shift from $\Delta V_{\mathrm{o},s}$-dominance at IDL~2 to $V_{\mathrm{c},s}$-dominance at IDL~3 generates a directional empirical prediction: analyst forecast dispersion should increase monotonically with IDL depth. Competitive-moat value contingent on full integration success, represented here by $V_{\mathrm{c},s}$, is inherently harder to estimate than $\Delta V_{\mathrm{o},s}$, which has observable operational proxies. This prediction is testable from analyst forecast data without access to the internal valuations that produced it. The exploratory AHP hypothesis runs in parallel: if pairwise-comparison consistency ratios increase with IDL depth, that constitutes independent evidence that the estimation problem is structurally harder at higher integration levels, beyond what individual rater heterogeneity alone would generate.

Single-rater probability estimates are vulnerable to anchoring and overconfidence, both well-documented patterns in expert elicitation. The AHP-based SRI protocol constrains expert judgement rather than replacing it. Pairwise comparisons expose internal inconsistency before aggregation, and the credentialed weighting scheme (\cref{eq:w-eff}) discounts estimates from raters with domain gaps or conflict-of-interest exposure. The resulting gain over current practice is narrow and concrete: disagreement can be assigned to a specific $p_s$ estimate for a specific option at a specific IDL stage, where a negotiated discount rate would otherwise absorb it.

$C_{\mathrm{r},s}$ enters the $V_s$ decomposition (\cref{eq:vs-decomp}) as a cost term, but regulatory compliance may also carry option value that the present framework does not model fully. Firms that complete EU AI Act compliance milestones early may acquire a market-access advantage that late compliers cannot replicate quickly; in regulated industries that advantage may persist. In the Swiss context, FINMA's 2024 guidance on AI-driven financial services \autocite{finma2024guidance08} and the Federal Council's sector-specific approach \autocite{bakom2025aiapproach} create distinct regulatory milestone paths whose option structure differs from the EU model. A fuller treatment of regulatory compliance as a staged option remains outside the present scope.

The lifecycle model (\cref{eq:v-lifecycle,tab:option-status}) and the synergy decomposition (\cref{eq:synergy}) reframe M\&A negotiation. Under current practice, acquirer and target disagree on a headline premium whose component drivers are opaque to both parties. The framework reduces this to a structured comparison of specific options, probabilities, and continuation values. An acquirer who assigns high probability to completing the IDL~3 transition can justify a higher premium transparently; a target that believes its $p_s$ estimates are conservative can argue that point at the option level rather than relitigating the discount rate. This does not eliminate negotiation difficulty; it replaces narrative disagreement with structured disagreement that due diligence can address.

Three additional finance-side gaps are worth making explicit before turning to the literature.

First, valuation multiples. EV/Revenue and P/E are the standard plausibility checks on any DCF, but they presuppose a homogeneous peer group. An IDL~3 firm with a platform-level data moat trades at fundamentally different multiples than an IDL~1 firm using off-the-shelf AI tools, even within the same sector and the same GICS sub-industry classification. Assigning a technology-sector median multiple to both produces a spurious comparison. The IDL taxonomy makes peer selection explicit and IDL-conditioned: the right comparable set is determined by integration depth, not sector label alone.

Second, terminal value. In a standard DCF, 60--80\% of total present value typically concentrates in the terminal-value term, a single growth rate applied to a normalised cash flow in perpetuity. For a firm at an IDL transition boundary, this growth rate implicitly bets on whether IDL~3 continuation value materialises. The lifecycle model (\cref{eq:v-lifecycle}) addresses this structurally: as milestones clear, option value migrates into the DCF baseline and the terminal-value calculation is applied to an already-adjusted cash flow rather than one that embeds unresolved option payoffs. The framework does not eliminate terminal-value sensitivity, but it confines it to the post-exercise DCF rather than allowing it to absorb milestone uncertainty silently.

Third, preferred equity and dilution. The pre-money headline valuation of an AI-native firm diverges from its common-equity value. Liquidation preferences, ESOP pools, and anti-dilution provisions systematically decouple the two: Gornall and Strebulaev find that US unicorn post-money valuations exceed fair common-equity value by 48\% on average due to the economic rights embedded in preferred shares \autocite{gornall2020squaring}. The Case Firm E12 illustrates this directly: the CHF~100 million pre-money is a Series seed headline. The $V_s$ decomposition operates on economic cash flows attributed to each option, which means dilution and liquidation-preference effects must be modelled separately in the capital-structure layer rather than embedded in either $p_s$ or $r_s$. Conflating them produces a valuation that is internally consistent at the option level but misleading at the equity level.

\Cref{tab:finance-gaps} maps each gap to its framework response; the common pattern is that the decomposition moves the problem to a layer where economic cash flows can be compared directly, rather than leaving it embedded in a multiple, a growth rate, or a headline number.

{\footnotesize
\begin{xltabular}{\textwidth}{p{3.0cm} X X}
\caption{Three conventional finance-side gaps and the framework's response.}
\label{tab:finance-gaps} \\
\toprule
Gap & Standard limitation & Framework response \\
\midrule
\endfirsthead
\multicolumn{3}{l}{\tablename~\thetable{} -- continued} \\
\toprule
Gap & Standard limitation & Framework response \\
\midrule
\endhead
\bottomrule
\endfoot
Valuation multiples & IDL heterogeneity makes peer groups non-comparable within a sector; median multiple mixes IDL~1 and IDL~3 firms & IDL classification conditions peer selection; multiples applied IDL-specifically rather than sector-wide \\[4pt]
Terminal value & 60--80\% of DCF value in a single perpetuity growth rate that implicitly bets on IDL~3 continuation & Lifecycle model explicitly migrates option value into DCF baseline as milestones clear; terminal value applied to adjusted post-exercise cash flow only \\[4pt]
Preferred equity / dilution & Pre-money headline $\neq$ common-equity value; liquidation preferences, ESOP pools, and anti-dilution decouple the two (average 48\% overstatement for US unicorns \autocite{gornall2020squaring}) & $V_s$ decomposition operates on economic cash flows; dilution and preference effects modelled separately in the capital-structure layer \\
\end{xltabular}
}

The Case Firm E12 functions as a demonstration, not as an empirical validation. The case was chosen for disclosure quality and for a clean AI-native structure, conditions that favour the framework's applicability and that bias case selection toward firms where the decomposition fits. One case establishes that the protocol can be applied coherently and yields a traceable band; it does not establish that the framework outperforms a conventional DCF on any observable metric, nor that the decomposition generalises to hybrid integrators, low-disclosure firms, or integration failures. A proper test requires a multi-case portfolio covering IDL levels, integration successes and failures, scored against realised post-exit valuations. That test is left to future research.

The framework connects three bodies of work. It introduces option structure into intangible-asset valuation \autocite{lev2001intangibles}, where the disclosure problem is recognised but no mechanism is provided for staging value realisation. It introduces integration-depth architecture into real-options applications in technology valuation \autocite{schwartz2000internet}, where uncertainty is modelled at the firm level without conditioning option structure on the depth of technology embedding. It also links AI adoption research \autocite{hansen2024aicmm,brynjolfsson2018jcurve} to valuation by connecting milestone achievement to firm value. The AHP component draws on established multi-criteria methodology \autocite{saaty1980ahp} and uses it for IDL-conditioned probability estimation in staged AI integration options.

\section{Conclusion}

The central claim of this paper is modest in scope but consequential in implication: standard valuation methods are not wrong about AI-integrated firms, but they are structurally incomplete. DCF, IDW~S~1's income approach, and market multiples all remain valid instruments for established cash-flow businesses. When applied to firms whose value is dominated by staged, contingent, and depth-dependent AI integration decisions, these methods compress option value, milestone risk, and continuation payoffs into parameters that obscure the very assumptions an analyst needs to challenge. The framework proposed here does not discard that machinery. It adds a decomposition layer, the IDL~0--3 taxonomy, the five-component $V_s$ structure, and the AHP-based SRI protocol, that makes the embedded assumptions explicit and individually auditable.

The framework's contribution lies in the informativeness of the disagreement it surfaces, beyond any improvement in point estimates relative to DCF or multiples. A conventional DCF applied to Case Firm E12 would yield a single present value whose sensitivity concentrates in the terminal growth rate and the discount rate, two parameters that absorb everything the analyst believes about AI integration without revealing what those beliefs are. The framework decomposes that belief into four milestone-level probabilities, five value components per milestone, and a gate structure whose failure conditions are contractually defined. Two analysts who disagree on the CHF~100 million pre-money can now locate their disagreement in $p_{\mathrm{M4}}$, in $V_{\mathrm{c},\mathrm{M4}}$, or in the 30\% discount rate. That precision is the contribution: a valuation whose internal structure can be interrogated.

Real-options theory has made similar promises before, and the track record warrants scepticism. Schwartz and Moon's option models for internet firms in 2000 were formally elegant but empirically uncalibrated; the biotech pipeline analogy that underpins much of staged-option pricing works well in pharmaceuticals, where phase-gate probabilities have decades of actuarial data, but lacks comparable base rates for AI integration milestones. The AHP-based probability architecture proposed here addresses part of this calibration gap. It replaces unstructured expert guesses with pairwise-comparison discipline, consistency checks, and credentialed aggregation; the resulting $p_s$ estimates still sit in the domain of expert judgement, since observed frequencies for AI integration milestones do not yet exist at scale. The framework's honesty about this point is deliberate: it surfaces where judgement enters the valuation, exposing what a discount rate would otherwise absorb. Whether that exposure leads to better decisions is an empirical question the framework itself cannot answer.

The Case Firm E12 illustrates both the promise and the boundary. On one hand, the framework produces a coherent valuation band (CHF~77m to CHF~123m) from transparent, citable inputs, and the concentration of risk in M4 ($|\delta w_s| = 11.7$, more than 60\% of total sensitivity) confirms the structural prediction that AI Provider valuations are dominated by late-stage continuation options. On the other hand, roughly half the firm's value rests on a single 24-month execution bet, a finding that no amount of decomposition makes less uncertain. The framework clarifies what the bet is and where it sits; it does not make the bet safer.

For European valuation practice, the implications extend beyond methodology. The measurement mismatch between how AI-integrated firms create value and how statutory standards capture it is not a temporary gap that market maturation will close. IDW~S~1's income approach, built for firms with demonstrable earnings histories, systematically underweights optionality; HGB's treatment of development costs ensures that balance sheets remain silent about the very investments that drive AI value. The framework proposed here operates on the economic layer beneath these reporting conventions: $C_{\mathrm{e},s}$ records AI development expenditure as economic cost regardless of its accounting classification, and the $V_s$ decomposition yields identical inputs under HGB, IFRS, and US~GAAP. If European venture capital is to close even part of the sevenfold valuation gap with U.S. peers at comparable development stages, the valuation infrastructure will need to accommodate option-loaded, milestone-gated firm profiles. This paper proposes one architecture for doing so.

\section{Limitations and Future Research}

Milestone identification depends on disclosure quality, and firms may reveal neither their integration depth nor the timing of intermediate gates with sufficient precision for external reconstruction. Parameter estimation for $p_s$, $V_s$, and $r_s$ remains sensitive to assumptions, especially at IDL~2--3 where comparables are scarce; the framework exposes that sensitivity rather than resolving it. IDL classification also requires judgement for hybrid firms operating at multiple levels across functions. The AHP component introduces further constraints: pairwise-comparison burden grows with criterion count, the IDL-specific calibration anchors remain illustrative priors rather than empirically derived base rates, and geometric-mean aggregation assumes a level of rater independence that may weaken when internal and external experts share the same organisational context. The credentialed weighting scheme partially addresses rater-quality and independence concerns, but assigning $w_i^{\mathrm{comp}}$ and $b_i$ is itself a judgement that must be documented and remains open to challenge. Option interdependency handling through attribution by assignment likewise depends on human judgement, because the interaction matrix is qualitative and does not replace formal coalition analysis. The case design is illustrative rather than inferential: a single firm with detailed milestone disclosure can test framework coherence, but not predictive accuracy. Case selection was also purposive: the Case Firm E12 offered rich investor disclosure, an AI-native business model, and a clean milestone architecture, conditions under which the framework is expected to fit well. Cases with weaker disclosure, hybrid integration across IDL levels, or post-hoc failed transitions remain outside the present evidence base, and generalisation requires a larger and more varied sample. Regulatory conditions are also moving targets, because the EU AI Act is not yet fully in force and the Swiss sector-specific approach is still evolving.

Future research should move in five directions.

The first direction is empirical validation. Applying the framework to a multi-case portfolio across IDL levels and integration outcomes, with option-level estimates scored against realised post-exit valuations, would test whether the decomposition improves on conventional alternatives rather than merely describing them.

The second direction is longitudinal. Tracking dynamic IDL transitions would show how value migrates from the option overlay into the DCF baseline as milestones clear over time.

The third direction is empirical calibration. Panel data on AI project outcomes across industries could anchor the SRI-to-$p_s$ mapping and refine milestone payoffs by sector.

The fourth direction is methodological. Bayesian updating for $p_{s+1 \mid s\;\text{cleared}}$, Shapley-value decomposition for option interdependencies, and robustness checks with alternative MCDM methods such as fuzzy AHP, ANP, or BWM would test how sensitive the framework is to its current simplifying choices.

The fifth direction is regulatory. As EU AI Act provisions take effect and Swiss guidance evolves, cross-jurisdiction evidence could show how compliance regimes alter milestone feasibility, risk, and continuation-option value.

\nocite{*}

\bigskip

\printbibliography

\swDeclarationsPage

\end{document}